\documentclass[12pt,pra,aps,amssymb,amsfonts,amsmath,tightenlines]{revtex4}

\usepackage[dvips]{graphicx}
\usepackage{dsfont, color} 

\newcommand{\re}{\mbox{$\rm e$}}
\newcommand{\rd}{\mbox{$\rm d$}}
\newcommand{\half}{\mbox{$\textstyle \frac{1}{2}$}}

\begin{document}

\title{Mathematical models for fake news\footnote{
To appear as Chapter 18 in  \emph{Financial Informatics: An Information-Based Approach to Asset Pricing}. D. C. Brody, L. P. Hughston \& A. Macrina (editors). Singapore: World Scientific Publishing Company (2022).}}
\author{Dorje C. Brody$^1$ and David M. Meier$^{2}$}

\affiliation{$^1$
Department of Mathematics, University of Surrey, Guildford GU2 7XH, UK, and \\
$^{2}$AXA Switzerland, General-Guisan-Strasse 40, 8400 Winterthur, Switzerland}

\date{\today}

\begin{abstract}
Over the past decade it has become evident that intentional disinformation in the political context -- so-called fake news -- is a danger to democracy. However, until now there has been no clear understanding of how to define fake news, much less how to model it. This paper addresses both of these issues. A definition of fake news is given, and two approaches for the modelling of fake news and its impact in elections and referendums are introduced. The first approach, based on the idea of a representative voter, is shown to be suitable for obtaining a qualitative understanding of phenomena associated with fake news at a macroscopic level. The second approach, based on the idea of an election microstructure, describes the collective behaviour of the electorate by modelling the preferences of individual voters. It is shown through a simulation study that the mere knowledge that fake news may be in circulation goes a long way towards mitigating the impact of fake news. 
\end{abstract}

\maketitle


\section{Introduction}
\label{sec:1}

``Our democracy is at risk'', summarizes the interim report published in July 2018 
by the UK House of Commons Digital, Culture, Media and Sport Committee (Collins 
\textit{et al}. 2018) on the dissemination of disinformation on social media for the 
purpose of manipulating the public in election and referendum voting. The prevalence 
of false stories on the internet has made it difficult for many to distinguish what is 
true from what is false.  The issue that lies at the 
heart of the current threat to the democratic process in the USA, the UK, and elsewhere 
is that, unlike in the physical sciences, in which the validity of a claim can be put to the test 
in a reproducible laboratory experiment, statements about past events are impossible 
to prove with the same degrees of scientific rigour. The existence of Holocaust denialists, for 
instance, illustrates how 
doubts about a major historical event can gain traction with certain individuals. To 
combat the forces of fake news it is important to view the issue through a scientific 
lens. Borrowing ideas from communication theory, the present paper aims at 
developing the mathematical theory that underlies the modelling of fake news.

In a broad sense the concept of ``fake news" has been around for centuries. In ancient 
China, for instance, the military strategist Kongming famously made use of 
state-sponsored 
disinformation to his advantage (Shou c. 290). During the Medieval period in 
Europe, the spreading of fake news often left violence and death in its wake (Soll 2016).  Once the technologies 
for mass printing had developed, fake news found a new kind of application in the 
form of sensationalist reporting to increase newspaper circulation. The demand for 
reliable information sources, however, was high in the twentieth century, especially in 
the post-war period, which made the running of a newspaper based on honest reporting 
a viable business model. This quasi-stable configuration of mainstream journalism based 
predominantly on honest reporting has been thrown out of its apparent equilibrium in the twenty-first century by the rapid growth of social media usage. As Soll 
(2016) puts it, ``It wasn't until the rise of web-generated news that our era's 
journalistic norms were seriously challenged, and fake news became a powerful force 
again.'' 

Advances in communication technology have their advantages and disadvantages. 
While it is undoubtedly true that the internet, for instance, has made it possible to 
access information that previously would have been difficult to acquire, it is also a matter of fact that the internet serves as a platform for the propagation of  irrelevant 
information, or ``noise''. As noise becomes more pervasive, it becomes increasingly 
difficult to access reliable information. Ultimately, as Norbert Wiener explained in his insightful 
book \textit{The Human Use of Human Beings} (Wiener 1954), one has to face the implications of the Second Law of 
Thermodynamics, which asserts that over time, noise will dominate (or, equivalently, 
in physicists' terminology, entropy will increase). Just as in any physical system where 
entropy can be reduced by means of external inputs (such as energy or force), to 
combat the domination of noise, concerted efforts have to be made, because noise will 
not disappear spontaneously. This has important implications for policy makers.

Today, fake news, fuelled by its speed of dissemination, has become a serious 
concern to society -- perhaps most importantly because it can endanger the 
democratic process. In response, academic research into various aspects of fake 
news has intensified recently, especially after the 2016 USA presidential election and 
the ``Brexit'' referendum in the UK on membership of the European Union. Broadly 
speaking, research carried out thus far has been primarily focused on the 
retrospective analysis of the impact of fake news (Allcott \& Gentzkow 
2017, Amador \textit{et al}. 2017, Bovet \& Makse 2018) and on the detection 
and prevention of fake news using deep learning and other related techniques 
(Conroy \textit{et al}. 2015, Shu \textit{et al}. 2017, Khajehnejad \& Hajimirza 
2018, Yang \textit{et al}. 2018). However, to 
address the issues surrounding the impact of fake news more generally, to make 
predictions of the impact of fake news, and to conduct a comprehensive scenario analysis, 
it is important that a consistent mathematical model should be developed that describes the 
phenomena resulting from the flow of fake news. For such a model to be useful, it should be intuitive (so that the model can be trusted as a 
plausible candidate) and tractable (so that model parameters can be calibrated against 
real data, and so that predictions can be made, either analytically or numerically). 
The 
purpose of this paper is to introduce a new framework for the mathematical 
modelling of fake news that fulfills these requirements. Our theory of fake news is straightforward to simulate and yet allows one to replicate 
qualitative features of empirical observations, as demonstrated below. 

\section{Fake news and communication theory}

Fake news is information that is inconsistent with factual reality. 
It is information that originates from the ``sender'' of 
fake news, is transmitted through a communication channel, and is then received, typically, by the general public. Hence any realistic model for fake news has to be built on 
the successful and well-established framework of communication theory. Indeed, 
this philosophy was already advocated by Wiener (1954), who wrote: 

\begin{quote} 
It is the thesis of this book that society can only be understood through a study 
of the messages and the communication facilities which belong to it; and that in the 
future development of these messages and communication facilities, messages 
between man and machines, between machines and man, and between machine and 
machine, are destined to play an ever-increasing part.
\end{quote} 

The modelling framework we propose in this paper embraces Wiener's philosophy. 
Specifically, we  shall apply and extend techniques of filtering theory -- a branch of 
communication theory that aims at filtering out noise in communication channels -- in 
a novel way to generate models that are well-suited for the treatment of fake news. 

The traditional applications of filtering theory are threefold: (i) extrapolation, or 
prediction, (ii) filtering, and (iii)  interpolation, or smoothing (Wiener 1949, 
Kailath 1974). 
Over the past decade, however, a fourth application of filtering theory has been 
developed, namely, in \textit{phenomenology} -- the description and modelling of 
observed phenomena. Perhaps surprisingly, the domain of applicability of the 
phenomenological use of filtering 
techniques ranges from scales as small as elementary particles and atoms  in physical 
systems (Brody \& Hughston 2006) to scales as large as human activities in social 
systems (Brody, Hughston  \& Macrina  2007, 2008). In the latter context, to describe the 
phenomena associated with social systems, it is particularly natural to employ the 
mathematics of filtering theory because the actions of an individual are ultimately 
based on the result of filtering the noisy information available to that individual. In 
other words, it is not the predictive power of filtering theory that is relevant; rather, 
it is the fact that the behaviour of people is guided by their predictions via filters, 
and thus \textit{this behaviour is itself susceptible to a filtering description}. These 
observations have opened  up a promising avenue towards new discoveries and 
novel applications, including the one that we explore here for the modelling of fake 
news.

The paper is organized as follows. We begin by explaining how the techniques of filtering 
can be applied in the context of behavioural phenomenology of an individual. We then 
apply this idea to the modelling of fake news as a modification of noise, and introduce 
our key assumption that, to a good approximation, people are rational inasmuch as 
they follow Bayesian logic in their decision making. Hence, in our approach, those who 
are influenced by fake news are not viewed as being irrational as such, but rather they 
lack the ability to detect and mitigate the changes caused by the presence of fake 
news in the structure of the noise they are exposed to. 
In going from the behavioural model of an individual to that of the electorate we are in 
effect introducing the idea of a ``representative voter" whose perception of the uncertain 
world represents the aggregation of the diverse views held by the public at large. 
We then examine the problem of estimating the release times of fake news, which in 
turn generates a new type of challenge in communication theory. This estimate is 
required for characterising a voter who is aware of the potential presence of fake 
news, but is unsure which precise items of information are fake. 
We show as an illustration the dynamics of the opinion-poll statistics in a referendum in the presence of a single piece of fake news. An application to an election in which multiple pieces 
of fake news are released at random times is then considered. Illustrative simulation results show 
that the qualitative behaviour of the dynamics of the opinion-poll statistics, seen for instance 
during the 2016 USA presidential election (Silver 2016), can be replicated by our 
model. 
To deepen the analysis further, we introduce what we call an ``election 
microstructure'' model in which we employ the same information-based scheme to 
describe the dynamical behaviour of individual voters and the resulting 
collective voting behaviour of the electorate under the influence of fake news. We 
conclude with a summary and a discussion of future directions.

\section{From communication theory to phenomenology}
\label{sec:2}

Let us begin by explaining the phenomenological application of filtering techniques. 
In our decision making we  are typically faced with uncertainties so that 
the most we can do is to arrive at a ``best guess'' for what the optimal decision might be. 
 Such situations are 
commonly encountered in our every-day lives. Suppose, for instance, one wants to travel from one location in a city to another. Should one take the bus or travel by underground? The walk to 
the bus stop may be shorter, but there might be heavy traffic; on the other hand, signal 
failures on the underground might result in a delay. From experience one has an initial view concerning whether travelling by bus or by underground is better. To formalise this notion, we let $p$ denote the 
\textit{a priori} probability that travelling by bus is the better choice. Correspondingly, $1-p$ 
will be the \textit{a priori} probability that it would be better to travel by underground. In 
other words, we have a binary random variable $X$ taking values, say, $(0,1)$, with corresponding 
probabilities $(p,1-p)$, where $X=0$ represents the bus being the better choice, and 
$X=1$ represents the underground being the better choice. 

The initial view, represented by the probability $p$, however, changes over time. A 
colleague who has been travelling on the underground might complain about 
the signal failure he encountered. The traffic news on the radio might suggest a 
delay on the bus route -- and so on. As time progresses, one's knowledge increases, but 
 uncertainties remain. We wish to model this type of dynamics. For this 
purpose, we assume for simplicity that reliable knowledge increases linearly in time, at a rate $\sigma$. 
The uncertainty, or noise, is modelled by a Brownian motion, denoted by 
$\{B_t\}_{t\geq0}$, 
which 
is assumed to be independent of $X$ because otherwise it cannot be viewed as 
representing pure noise. Hence, the flow of information, which we denote by 
the time series $\{\xi_t\}_{t\geq0}$, 
can be expressed in the form 
\begin{eqnarray}
\xi_t = \sigma X t + B_t . 
\label{eq:1} 
\end{eqnarray} 
The quantity of interest is the actual value of $X$.  However, since there are two unknowns, 
$X$ and $\{B_t\}$, and only one known, $\{\xi_t\}$, a rational individual will consider the 
probability that $X=0$ (or $X=1$) conditional on the information contained in the time 
series $\{\xi_s\}_{0\leq s\leq t}$ gathered up to time $t$. In other words, writing 
$(x_0,x_1)=(0,1)$, one considers the conditional probability 
${\mathbb P}(X=x_i|\{\xi_s\}_{0\leq s\leq t})$. In 
this simple model the time series $\{\xi_t\}$ is a Markov process, from which it follows that the 
conditional probability equals ${\mathbb P}(X=x_i|\xi_t)$. 

The logical step of converting the prior probabilities ${\mathbb P}(X=x_i)$ into 
the posterior probabilities ${\mathbb P}(X=x_i|\xi_t)$ is captured by the Bayes formula: 
\begin{eqnarray}
{\mathbb P}(X=x_i|\xi_t) &=&  \frac{{\mathbb P}(X=x_i)\rho
(\xi_t|X=x_i)}{\sum_{j} {\mathbb P} (X=x_j)
\rho(\xi_t|X=x_j)} . 
\label{eq:c2.33}
\end{eqnarray}
Here the conditional density function $\rho(\xi_t|X=x_i)$ for the random variable 
$\xi_t$ is defined by the relation
\begin{eqnarray}
{\mathbb P}\left(\xi_t\leq y|X=x_i\right)=\int_{-\infty}^y
\rho(\xi|X=x_i)\,\rd\xi, 
\label{eq:c2.34}
\end{eqnarray}
and is given by
\begin{eqnarray}
\rho(\xi|X=x_i)=\frac{1}{\sqrt{2\pi t}} \exp\left(-
\frac{(\xi-\sigma x_it)^2}{2t}\right). 
\label{eq:c2.35}
\end{eqnarray}
This follows from the fact that, conditional on $X=x_i$, the random variable $\xi_t$ is 
normally distributed with mean $\sigma x_i t$ and variance $t$. Recalling that $(x_0,x_1)=(0,1)$ we thus obtain
\begin{eqnarray}
{\mathbb P}(X=x_i|\xi_t) =\frac{p_i\exp\left( \sigma x_i \xi_t-
\frac{1}{2} \sigma^2 x_i^2 t\right)} {p_0 + p_1 
\exp\left( \sigma \xi_t-\frac{1}{2} \sigma^2 t\right)} , 
\label{eq:5}
\end{eqnarray}
where $p_0=p$ and $p_1=1-p$. Inferences based on the use of (\ref{eq:5}) are 
optimal in the sense that they minimize the uncertainty concerning the value of 
$X$, as measured by the variance or entropic measures subject to the information 
available. Hence, a rational individual will at any given time act in accordance with 
the changing views expressed in (\ref{eq:5}).  In the example mentioned above, for 
instance, the option to travel by bus would be chosen by a rational individual if at 
the time $t$ of departure it holds that ${\mathbb P}(X=0|\xi_t) > \half$. There 
are suggestions that people need not act rationally as anticipated by the Bayes 
rule (Kahneman \& Tversky 1974, Grether \& Plott 1979), but other 
studies suggest that the Bayes logic is nevertheless a dominant factor (El-Gamal 
\& Grether 1995). It is our opinion that in the context of signal processing, given 
the prior, it is reasonable to assume that people intuitively follow a Bayesian line of 
thinking. 

The mathematical framework outlined above is that of nonlinear filtering, familiar from communication theory. In communication theory, the random variable $X$ in the 
first term of \eqref{eq:1} represents the ``signal'' that one wishes to estimate in the 
presence of ambient noise represented by $\{B_t\}$. The parameter $\sigma$ then 
determines the signal-to-noise ratio. More generally, the signal 
typically changes in time, which can be represented by a time series 
$\{X_t\}_{t\geq0}$. The 
case of a fixed $X$ considered here can thus be viewed as a special case, which was 
studied by Wonham (1965).  It is important to emphasise, however, that in the context 
of communication theory, there is a ``sender'' actively transmitting the signal; whereas 
in our behavioural analysis, we often encounter circumstances where there  are 
receivers but no senders of the signal, because the random variable $X$ may represent 
the outcome of a future real-world event that is not known to anyone, and hence cannot 
be transmitted by anyone. 
Nevertheless, $X$ does exist, and people make 
decisions in accordance with their best estimate about $X$ based on partial information 
available to them. 
This is the sense in which mathematical techniques in communication theory can be 
applied to describe observed phenomena in science and in society. 
In what follows we shall extend the foregoing 
ideas to tackle the problem of modelling fake news and its impact. 

Before we proceed, we remark, incidentally, that although we have considered here the 
situation involving two possible alternatives, represented by the binary random variable 
$X$, the complexity of the analysis remains largely unchanged when $X$ can take multiple values. 
Likewise, the assumption that knowledge concerning the value of $X$ is revealed at a 
constant rate $\sigma$ can be relaxed without affecting analytical tractability 
(Wonham 
1965). In this case, the first term in (\ref{eq:1}) is replaced by $X\int_0^t \sigma(s) \rd s$, 
where $\sigma(s)$ represents the information flow rate at time $s$. Circumstances where the  
 value of $X$ is revealed with certainty over a finite time horizon can be modelled by replacing the Brownian noise $\{B_t\}$ in (\ref{eq:1}) 
with a Brownian bridge over that time horizon (Brody, Hughston \& Macrina 2007). More generally, 
in situations where Brownian motion is not an appropriate model for uncertainties 
(if the noise process can have jumps, for instance), then one can model the noise term by a L\'evy 
process, again without affecting analytical tractability (Brody \& Hughston 2013).

\section{Modelling fake news}
\label{sec:x1}

When fake news is released, for 
instance by a malicious individual aiming to mislead the public, the false 
information is superimposed on other information. 
However, fake news, by its nature, does not represent truth statements about 
the value of the quantity 
$X$ that people wish to determine, so it cannot be viewed as forming part 
of the signal that helps people discover the true value of $X$. On the other hand, 
from the point of view of signal processing, anything that is not part of the signal 
can be viewed as noise.  
Following this logic, we thus arrive 
at our model for the information process in the presence of fake news: 
\begin{eqnarray}
\eta_t = \sigma X t + B_t + F_t , 
\label{eq:6} 
\end{eqnarray}
where the time series $\{F_t\}_{t\geq0}$ represents fake news. 
The noise term $\{B_t\}$, which has no bias, 
represents the aggregation of a large number of unsubstantiated rumours and 
speculations about the value of $X$. The central limit theorem then 
suggests the normality of the noise distribution, making Brownian motion a viable 
candidate for modelling noise. The time series $\{F_t\}$ thus introduces an additional 
bias.
We can now offer a precise mathematical answer to the open issue 
raised in the UK House of Commons Committee Report on fake 
news (Collins, \textit{et al}. 2018):
\begin{quote} 
There is no agreed definition of the term `fake news', which became widely used 
in 2016. Claire Wardle . . . told us . . . that ``when we are talking about this huge 
spectrum, we cannot start thinking about regulation, and we cannot start talking 
about interventions, if we are not clear about what we mean''.
\end{quote} 
With this in mind, we propose the following: \vspace{2mm}
\newline
\textbf{Definition} (Fake News). \textit{A time series $\{F_t\}$ 
appearing in the information process {\rm (\ref{eq:6})} represents ``fake news'' if 
it has a bias so that ${\mathbb E}[F_t]\neq0$, where ${\mathbb E}$ denotes 
expectation}. 
\vspace{2mm}
\newline 
The existence of bias here is important, for otherwise $\{F_t\}$ would merely 
represent further noise, 
rather than deliberate misinformation. It is certainly the case that additional unbiased 
noise is a nuisance, delaying the process of discovering the truth; but it cannot ultimately 
drive the public away from discovering the truth. 
That said, in some circumstances there are advantages in merely delaying 
the process of truths being uncovered, in which case a release of an 
unbiased noise with ${\mathbb E}[F_t]=0$ would suffice, and one could 
describe this situation as representing a mild form of disinformation. Such a 
scenario, however, is in effect equivalent to manipulating the information 
flow rate $\sigma(t)$, and corresponds to the 
disinformation model proposed in Brody \& Law (2015). 
As regards the statistical dependency 
between $\{F_t\}$ and $X$, there are two situations that can arise: one in which no one 
knows the value of $X$, in which case $\{F_t\}$ should be independent of $X$, and one in 
which the value of $X$ is known to a small number of individuals who may wish to 
disseminate fake news, in which case $\{F_t\}$ may well be dependent on $X$.

The idea that information-based models of the kind represented in (\ref{eq:1}) can 
be extended to model deliberate misspecifications of the truth has previously been 
envisaged (Brody \& Law 2015). The proposal there was that a malicious individual 
who wishes to manipulate the public can alter the value of the information flow rate 
$\sigma$. Hence the public would make their inferences based on a particular value of 
$\sigma$, whereas the actual value of $\sigma$ is in fact different, and as a 
consequence the public is misled. Such a scheme amounts to setting $F_t=\mu Xt$ 
for some $\mu$, which might arise in an election microstructure model described below in which the value 
of $X$ may be known to the candidate but not to the public, thus allowing the candidate 
to transmit $X$-dependent fake news. More generally, taking into account the randomness 
in the release time, one might consider a fake-news structure of the form 
$F_t=\mu X (t-\tau){\mathds 1}\{t>\tau\}$. 
Here ${\mathds 1}$ 
denotes the indicator function so that ${\mathds 1}\{A\}=1$ if $A$ is true and 
${\mathds 1}\{A\}=0$ otherwise. 
This is equivalent to having the information process $\xi_t={\hat\sigma} X t + B_t$ 
with a random ${\hat \sigma}$, for which analytic expressions for the conditional 
probabilities can be obtained (Brody \& Law 2015).

In order to analyse the effects of fake news, it will be useful to classify members of the 
public into three categories. We define 
Category I to indicate those who are unaware of the potential existence of fake elements 
in the information they see. Nevertheless, they act rationally in the sense 
that they make their estimates in 
accordance with formula (\ref{eq:5}), except that $\eta_t$ is substituted in place of 
$\xi_t$. In other words, they ``correctly'' infer the posterior probability, but based on the mistaken 
belief that the information they are receiving is of the type (\ref{eq:1}), while in reality it 
is of the type (\ref{eq:6}). As we shall see, the people in this category are most vulnerable to exposure to 
fake news. We denote by Category II those members of the public who are aware of the potential existence of 
fake news, but without knowing precisely the times at which the items of fake news in the time series $\{F_t\}$ are released. These individuals face the most technically challenging task, because in their estimation they must deal with  
 three 
unknowns, $X$, $\{B_t\}$, and $\{F_t\}$, but only one known, $\{\eta_t\}$. 
As we shall see, although analytic expressions for the conditional probability 
${\mathbb P}(X=x_i|
\{\eta_s\}_{0\leq s\leq t})$ can be obtained,  
the analysis is rather more involved than the one for Category I voters. 
Thus, the people in this category are considerably more cognisant of the uncertainties in their estimates than 
those in Category I. Finally, Category III consists of those people that are highly 
informed, to the extent that they know the 
values of the time series $\{F_t\}$. Because $\{F_t\}$ contains no information relevant 
to $X$, they can simply disregard $\{F_t\}$ from their information $\{\eta_t\}$ and use $\xi_t = 
\eta_t - F_t$ instead to work out their posterior belief according to (\ref{eq:5}). Like those in 
Category I, people in Category III would tend to be assertive in their judgements. 
We note however that a Category-III individual should be regarded to some 
extent as an idealization. After all, it is an almost insurmountable task for an individual to identify perfectly which items of news are fake and which ones are not.

\section{Estimating the arrival times of fake news}
\label{sec:x2}

From the point of view of those belonging to Category II of our classification, there are two issues to address: first, one must estimate whether the information source has 
been contaminated with fake news. Based on this consideration, one must then determine the conditional probability 
${\mathbb P}(X=x_i|\{\eta_s\}_{0\leq s\leq t})$, which gives the best estimate for 
the likelihood of the event $X=x_i$. Note that the former issue 
amounts to working out the conditional probability $f_t(u) \rd u = 
{\mathbb P}(\tau\in\rd u|\{\eta_s\}_{0\leq s\leq t})$ that $\tau$ takes a value within 
the small interval $[u,u+\rd u]$, which we shall calculate. 

Before we proceed, we remark that there is an extensively-studied research area within
communication theory that is concerned with ``change-point detection'' or ``disorder detection'' (Page 1954, 
Shiryaev 1963a, 1963b). The nature of this problem is as follows. One observes a time 
series with the property that the structure of the series changes at some random time.  In a situation where this transition is not immediately apparent from observed data the task is to detect whether a ``regime change'' has occurred, and if so, when it might 
 have occurred. 
Stated mathematically, a prototype of such a problem is to detect the random time 
$\tau$ at which a Brownian motion acquires a drift (Karatzas 2003), or to detect the 
random time $\tau$ at which the jump rate of a Poisson process changes from one to another 
(Galchuk \& Rozovskii 1971, Davis 1976, Peskir \& Shiryaev 2002). In the Brownian context, therefore, the observation is modelled by a process of the 
form $B_t+\mu(t-\tau) \, {\mathds 1}\{t\geq\tau\}$, which is the same as setting 
$\sigma=0$ and $F_t=\mu(t-\tau) \, {\mathds 1}\{t\geq\tau\}$ 
in our information process $\{\eta_t\}$.  
Thus, we see that the analysis of fake news introduces 
a new type of problem in signal detection, where one wishes to detect the unknown 
signal $X$ in a noisy environment where the structure of the noise switches from one 
regime to another at a random time. In other words, one is trying to detect the moment 
of the switch -- that is, the time $\tau$ at which fake news emerges -- while at the same 
time estimating $X$. It 
should be intuitively clear that such a problem will have a wide range of applications 
beyond fake news analysis. 

With this in mind, let us examine the conditional density for $\tau$ in the context of at most one piece of fake news. For this purpose, 
let us consider the model for fake news given by 
$F_t=m(t-\tau) \, {\mathds 1}\{t\geq\tau\}$, where 
$m:{\mathds R}\to{\mathds R}$ is an arbitrary function which we assume 
to be at least once differentiable. This is the form of the fake news model that a Category II voter {\em assumes} to be the true model. Then a calculation shows that the conditional density for $\tau$ is given by 
\begin{eqnarray} \label{est_CatII}
f_t(u) = \frac{f_0(u) \sum\limits_{i} p_i \, \re^{ 
\int_0^t (\sigma x_i + m'(s-u){\mathds 1}\{s\geq u\}){\rm d}\eta_s - \frac{1}{2} 
\int_0^t (\sigma x_i + m'(s-u){\mathds 1}\{s\geq u\})^2 {\rm d} s }}
{\int_0^\infty f_0(w) \sum\limits_{i} p_i \, \re^{ 
\int_0^t (\sigma x_i + m'(s-w){\mathds 1}\{s\geq w\}){\rm d}\eta_s - \frac{1}{2} 
\int_0^t (\sigma x_i + m'(s-w){\mathds 1}\{s\geq w\})^2 {\rm d} s } \rd w }, 
\nonumber \\ 
\end{eqnarray}
where $m'(u)=\rd m(u)/\rd u$ and  $p_i = \mathbb{P}(X = x_i)$. Note that $f_0(u)$ 
is the \textit{a priori} density for $\tau$, which reflects the initial view on how the release timing 
is distributed. Hence, the best estimate for the release time of fake news is given by 
$\int_0^\infty u f_t(u) \rd u$. 

With the help of $f_t(u)$ we are now in a position to address the estimation 
of $X$ for the Category II public. From the tower property of conditional 
expectation we can write 
\begin{eqnarray}
{\mathbb E}[X|\{\eta_s\}_{0\leq s\leq t}] = 
{\mathbb E}[{\mathbb E}[X|\{\eta_s\}_{0\leq s\leq t},\tau]|
\{\eta_s\}_{0\leq s\leq t}].
\end{eqnarray}
Our strategy is to work out the inner expectation first. Since conditional on $\tau$ 
the information process $\{\eta_t\}$ is Markov, this reduces to 
$Y(\eta_t,\tau)={\mathbb E}[X|\eta_t,\tau]$, which, on account of the Bayes 
formula, is given by 
\begin{eqnarray}
Y(\eta_t,\tau) = \frac{ \sum_k x_k p_k \exp\left( \sigma x_k 
\eta_t - \frac{1}{2} \sigma^2 x_k^2 t - \sigma x_k m(t-\tau) \mathds{1}
\{t\geq\tau\} \right)}{ \sum_k p_k  \exp\left( \sigma x_k \eta_t - 
\frac{1}{2} \sigma^2 x_k^2 t - \sigma x_k m(t-\tau) 
{\mathds 1}(t\geq\tau) \right)} . 
\end{eqnarray}
It follows that the best estimate for $X$ is given by 
$\int_0^\infty Y(\eta_t,u) f_t(u) 
\rd u$. 
The foregoing analysis shows that although it is possible to deduce a closed-form 
expression for the best estimate of the quantity $X$ that one wishes to 
determine, the procedure is somewhat intricate.  Note that we have treated the case 
of a single piece of fake news being released at a random time. However, the 
mathematical formalism describing Category II voters can be extended in a 
straightforward manner to the case where multiple items of fake news are released. 
We shall see below simulations of this more general setup. 

\section{Representative voter framework}
\label{sec:3}

With the foregoing discussion in mind, the classification arising from our modelling 
setup is clear: there are those members of the public who are easily misled and 
manipulated since they ignore, by choice or otherwise, the possible existence of 
fake news; those who are wary of the potential existence of fake news but cannot 
be too certain about their judgements; and those who are able to detect and 
disregard fake news. The characteristics of these three categories can be captured 
most easily by means of simulation studies. For this purpose, let us consider a 
yes-or-no referendum scenario with a simple linear model for fake news: 
$F_t = \mu ( t-\tau) \, {\mathds 1}\{t\geq\tau\}$. 
Here, $\mu$ is a constant whose sign corresponds to which way the public will be 
misled, and $\tau$ is an exponentially distributed random variable, with density 
$f(u)=\lambda\re^{-\lambda u}$, $\lambda>0$ a constant. 

We let the realization $X=1$ be associated with the ``yes'' vote and $X=0$ with 
the ``no'' vote. 
In particular, we interpret the likelihoods for the events $X=1$ 
and $X=0$ to represent the aggregate opinion of the public. That is to say, 
$p={\mathbb P}(X=1)$ represents the current percentage of the public who 
intend to vote ``yes'', and conversely for $1-p={\mathbb P}(X=0)$. Thus the 
\textit{a priori} probability $p$ can be calibrated from today's opinion-poll 
statistics. The public opinion, however, changes over time in accordance with the 
flow of information, and hence the \textit{a priori} probability will be updated 
accordingly. 

It is worth remarking that whereas in the earlier discussion on the phenomenological 
application of filtering techniques we described the modelling of the 
behaviour of an individual, here we take an ensemble point of view in which the 
\textit{a priori} probability $p$ refers to the aggregation of the diverse opinions 
held by the public (see Brody \& Hughston 2013 for a similar concept). The 
idea that we advocate here is that of a ``representative voter'', 
in analogy with representative agent models in economics. This is useful for the 
purpose of obtaining a qualitative understanding of the observed phenomena. We shall later 
in the paper return to the level of individual behaviour by introducing an alternative ``election microstructure'' framework in 
analogy with market microstructure models in economics. This is useful for 
the purpose of developing countermeasures against fake news, and for policy 
making. 

In the ensemble formalism, the event that the random variable $X$ takes the 
value zero can be interpreted as the situation in which the totality of 
voters agree that they should be voting for the `no' outcome. 
If the information flow-rate parameter $\sigma$ were constant, then 
indeed the value of $X$ would be revealed asymptotically, and the public opinion would 
thus eventually converge to one choice or the other, over an infinite time horizon. Of 
course, this rarely happens in real life, because $\sigma$ will be 
time dependent and 
tends to vanish after the election has taken place. Nevertheless, it is not unreasonable in the present context to make the assumption that $\sigma$ is a constant, since we are only interested in the dynamics up to the polling day. 

\begin{figure}[t!]
\centerline{
\includegraphics[width=0.50\textwidth]{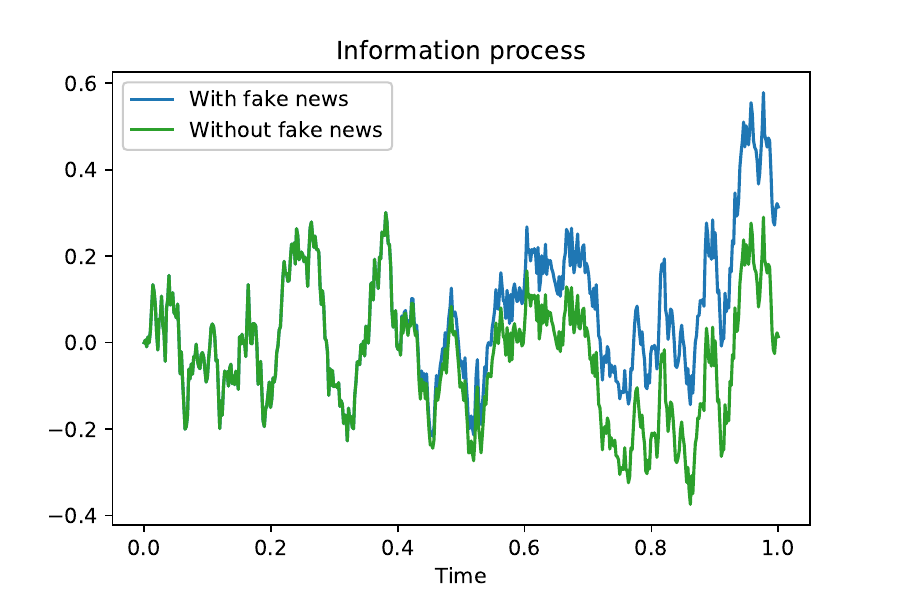}
\includegraphics[width=0.50\textwidth]{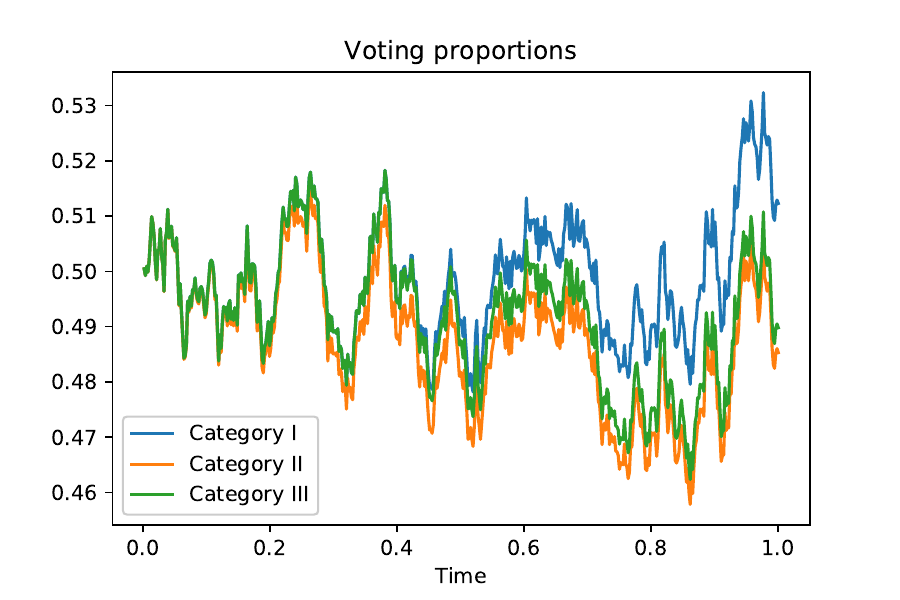}
}
\caption{\footnotesize{
\textit{Referendum simulation}. 
A typical sample path associated with the linear fake-news model is sketched. 
The left panel shows the information processes $\{\xi_t\}$ without fake news 
and $\{\eta_t\}$ with fake news, released here at time $0.4$. 
The evolution of public opinion over time is shown on the right panel for the 
three types of voters: (I) those who are unaware of the existence of fake news; 
(II) those who are aware of the potential existence of fake news; and (III) 
those who are able to disregard fake news fully. 
Here, $X=0$ corresponds 
to the ``no'' outcome, and $X=1$ corresponds to the ``yes'' outcome. We see that in this 
particular sample path, a majority of Category III voters will vote for the ``no'' outcome. However, it is 
evident that a piece of fake news that attempts to sway the voters in favour of the 
``yes'' outcome has had the effect of moving the percentage figure of Category I voters by 
a little over 2\%, just enough to change the majority vote to ``yes''.  
On the other hand, the relative proximity of Category II and Category III curves shows that Category II voters are able to largely correct for their exposure to fake news. The parameters chosen for 
the simulation are $\mu=0.5$, $\sigma=0.3$, $\lambda=3$, $p=0.5$, and for the random 
variable $X$, the sample outcome is $X=0$.  
}}
\label{label_here}
\end{figure}

With these preliminary remarks in mind, we have simulated typical sample paths 
corresponding to the three voter categories for various choices of the 
model parameters $(\mu,\sigma,\lambda,p)$.  The results of the simulation are illustrated in 
Figure~\ref{label_here}. 
The expected main feature of the simulation paths is that 
the voters unaware of the existence of fake news are indeed swayed in the 
intended direction. What is perhaps less obvious is that the Category II 
voters, who are aware of the possible existence of fake news, but do not know the timing of its release, tend to overcompensate for the possibility that the information they are 
receiving may be contaminated. As a consequence, their estimates deviate away from the 
``correct'' (Category III) estimate before fake news is released. However, once the fake news is released, Category II voters do surprisingly well at removing the effects of fake news from their estimates. One can interpret this as an indication that mere knowledge of the possibility of fake news is already a powerful antidote to its effects. Further evidence of this will be seen below in our discussion of election microstructure models.

\section{Application to opinion-poll statistics in an election}
\label{sec:4}

We proceed to model the dynamics for opinion-poll statistics in an election 
where fake news is present. For simplicity, we shall assume that there are 
two dominant candidates, represented by a binary random variable $X$, taking 
the values $0$ and $1$ that correspond to the two candidates, with 
\textit{a priori} probabilities $p$  and $1-p$, respectively. 
As indicated earlier, in the representative voter framework the \textit{a priori} 
probability $p$ represents the diverse opinions and mixed views initially held by the 
public about which value $X$ should take. The opinion of the public, 
however, evolves over time in accordance with the revelation of information, which 
we model by the process $\{\eta_t\}$. In the event in which the public at large 
favours candidate $0$, a malicious 
individual supporting candidate $1$ might in response decide to release a 
false statement about candidate $0$. According to our previous model 
$F_t = \mu ( t-\tau) \, {\mathds 1}\{t\geq\tau\}$ 
the contribution of fake news continues to grow at a rate $\mu>0$. 
However, in reality one might expect that over time the strength of any one fake 
news item diminishes. 
Thus, we shall consider here a 
modification of $F_t$ whereby the strength of fake news initially grows 
linearly in time, but is then damped exponentially. In other words, we let the contribution to $F_t$ of a single item of fake news released at time $\tau$ be given by $m ( t-\tau) \, {\mathds 1}\{t\geq\tau\} $, 
where $m(u) = \mu u \re^{-\alpha u}$ for some damping rate $\alpha>0$, 
and $\tau$ is distributed according to $f(u)=\lambda\re^{-\lambda u}$. Once 
the effects of fake news are sufficiently damped, the public support may 
revert back towards the direction of candidate $0$. However, another item of
fake news may be released, and so on. The process will be repeated until   
polling day. 

\begin{figure}[t!]
\centerline{
\includegraphics[width=0.50\textwidth]{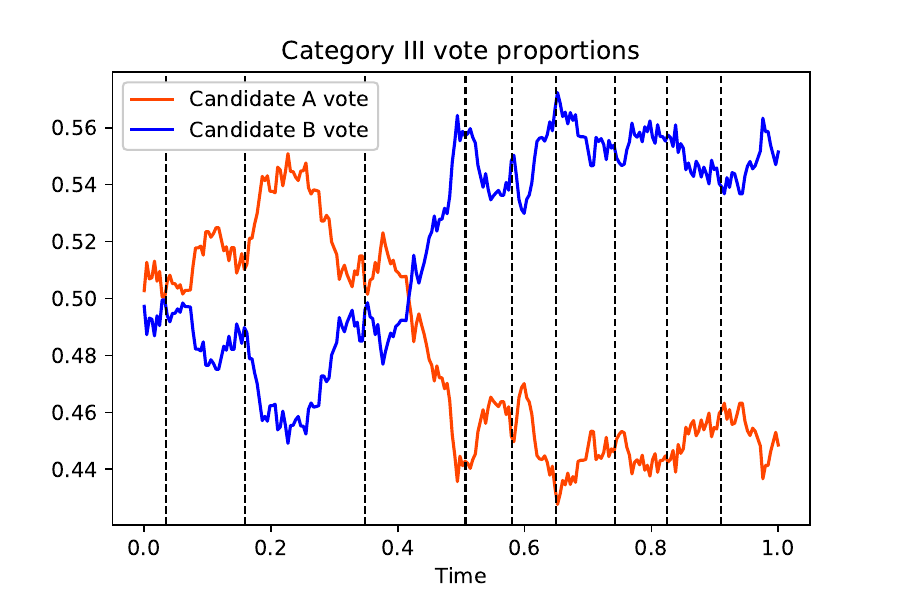}
\includegraphics[width=0.50\textwidth]{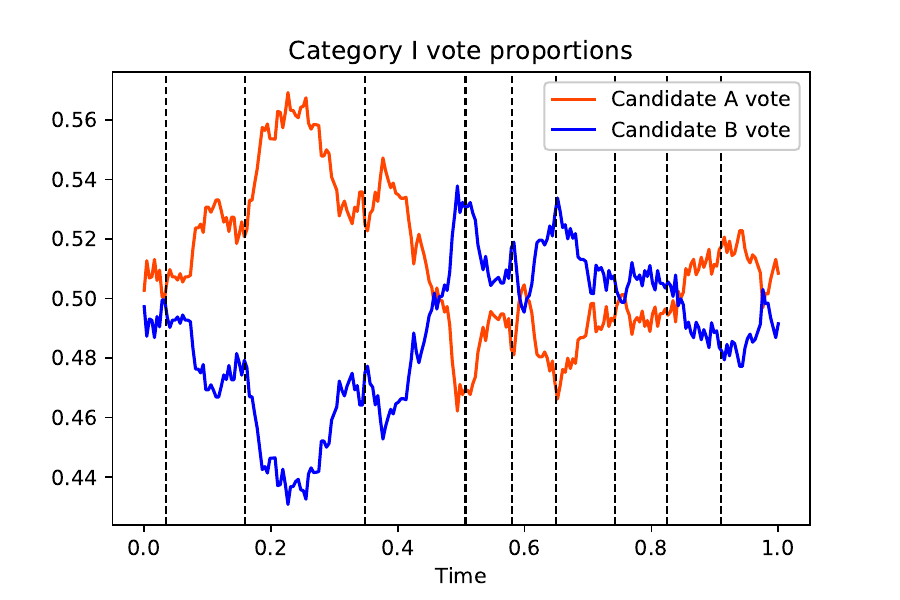}
}
\caption{\footnotesize{
\textit{Election dynamics without and with fake news}. 
In this example, in the absence of fake news Candidate B, who initially 
struggles somewhat in the opinion poll, nevertheless manages to win the election 
comfortably (left panel). However, with persistent fake news (released in this example on nine occasions, indicated by the vertical lines), the public is misled 
sufficiently for Candidate A to secure a narrow victory (right panel). Of the 
100K simulations we conducted with the same parameter choice, the losing candidate 
ended up winning the election owing to fake news in some 30\% of the cases.  
The parameters are as follows: $\sigma=0.3$, $\mu=2$, $\alpha=5$, $p=0.5$, 
and $\lambda=10$. 
}}
\label{fig2}
\end{figure}

The structure of our model should now be evident. At time $\tau_1$ comes the 
release of the initial piece of fake news, to be damped gradually, at time $\tau_2$ 
comes the next release, and so on. The waiting times between fake news releases are 
modelled by an exponential distribution. In other words, the release times are the jump 
times of a Poisson process. We are interested in a  simulation study of the dynamical 
evolution of opinion-poll statistics in the presence of various pieces of fake news. 
For simplicity, we consider fake news to be one sided, i.e. it is released only by the 
supporters of one of the candidates. For the purpose of simulation we shall also 
assume that the damping rates for the various fake news releases are all 
equal, and similarly for the linear growth rates. Typical sample paths resulting from the 
simulation studies are shown in Figure~\ref{fig2}.  For clarity, we show only Category 
I and Category III voters. In this way, we can see the full, unmitigated effect of fake 
news. We see that for this sample path the candidate who would win the election in 
the absence of fake news ends up losing it. In fact, based on the parameter choice 
indicated in the figure caption, we found that the likelihood of the ``losing'' candidate 
(who would have lost the election in the absence of fake news) ending up 
winning is about 30\%. This number, of course, depends on the choice of model 
parameters, and thus can be increased arbitrarily by increasing the fake news strength 
or frequency parameters. Thus, while each element 
of fake news is damped exponentially in time, a persistent attack on democracy can 
and will succeed if no action is taken against it.

\section{Election microstructure models}\label{Sec:ElMic}

The representative voter framework presented above is highly effective in modelling 
the  stylized aspects of fake news and its impact at a phenomenological level. It can 
also be applied in an efficient scenario analysis by means of simulation studies, 
for instance, the parameter dependence of the likelihood of fake news changing the outcome of an election. 
We shall 
now turn to an alternative formulation that focuses on the microscopic level of individual 
voters to deduce the macroscopic behaviour of the general public. While the mathematical ingredients used in this election microstructure model are essentially the 
same as those used for the representative voter framework, there is one important 
conceptual difference, namely, that in the election microstructure model the signal in the 
information process can be transmitted by a sender (e.g., the candidate). What follows is 
framed in the language of political elections, but the treatment is equally 
applicable to referendums.  

Our model is based on the following idea, which represents a somewhat simplified 
characterization of the mechanism by which an individual arrives at their preferred 
choice of candidate: when deciding who to vote for in an election, an individual will 
attempt to evaluate candidates with respect to a number of issues, or factors. For 
instance, what are the views of the candidates 
on taxation, on social welfare, on 
freedom of the press, on immigration, on abortion, on transport, on gun control, on 
healthcare, on public spending, and so on? Nonpolitical elements may also be considered, 
such as the state of health of the candidate, or the level of personal integrity, etc. 
As we shall outline below, the positions of the candidates on these factors are then transformed into an overall score, and 
the voter ultimately picks the candidate with the highest score. 

To proceed let us assume that there are $K$ independent such factors in an election 
with $L$ candidates and $N$ voters. We let $X_k^l$ denote the 
$l$-th candidate's position on the $k$-th factor. 
However, the positions of the candidates on these factors, if they were in office, are not 
always transparent to voters. Nevertheless, during an election 
campaign, candidates will attempt to communicate their positions to the electorate. 
This communication flows through a number of channels: through advertising by the 
campaigns, through publications by news outlets, through word-of-mouth, through 
social media, and so on. The members of the electorate thus obtain partial information, 
which we shall model as before in the form 
\begin{eqnarray}
\eta_t^{k,l} = \sigma_{k,l} X_k^l \, t + B_t^{k,l} + F_t^{k,l} 
\end{eqnarray} 
for $k=1, 2, \ldots, K$ and $l=1, 2, \ldots, L$. In other words, we have a total of $KL$ 
information processes $\{\eta_t^{k,l}\}$. Some of these may be contaminated by fake 
news, which can take different forms. For instance, candidates may try to 
make themselves look more palatable to a wider section of the electorate than they 
perhaps are (such as by suppressing about their health, or by downplaying their views on taxation). They 
may run negative ads to spread untruths or half-truths about their 
opponents. Or there may be malicious agents spreading entirely fabricated stories 
supporting one candidate or the other. It is this latter type of fake news that can 
spread quickly on social media and has thus attracted much attention recently. 

A member of the electorate has their own views on how attractive a candidate with a 
given set of factor values is. We model this by a {\em score function}, which can differ 
between individuals and assigns a measure of attractiveness to a candidate with a 
given factor vector ${\boldsymbol X}^l$. Hence, a member of the electorate will vote 
for the candidate with the highest score. For the present paper, we 
consider linear score functions. Thus, the score $S_n^l$ of candidate $l$ determined 
by voter $n$ is given by 
\begin{eqnarray}
S_n^l = \sum_{k=1}^K w_{n}^k X_k^l = {\boldsymbol w}_n \cdot 
{\boldsymbol X}^l,
\end{eqnarray} 
supposing the vector ${\boldsymbol X}^l$ were known. The weights $\{w_n^k\}$ may be 
positive or negative, with magnitudes that may be large or small depending on how 
strongly the voter feels about a given issue. Keeping in mind that the factors need 
to be estimated from the available information, voter $n$ will choose candidate $l$ 
over candidate $l'$ if and only if
\begin{eqnarray}
\sum_{k=1}^K w_n^k \left( {\mathbb E}_t[X_k^{l}]-{\mathbb E}_t[X_k^{l'}]\right) > 0, 
\label{decision}
\end{eqnarray}
or more succinctly ${\boldsymbol w}_n \cdot( {\mathbb E}_t[{\boldsymbol X}^{l}] 
- {\mathbb E}_t[{\boldsymbol X}^{l'}]) > 0$. 
Here we let ${\mathbb E}_t[-]$ denote expectation conditional on having observed the information processes $\{\eta_s^{k,l}\}$ up to time $t$. 
Each component of the conditional expectations $ {\mathbb E}_t[{\boldsymbol X}^{l}]$ can be worked out by following the methodology described earlier, on account of the independence of policy factors $X_k^l$ 
and ${X}_{k'}^{l}$ for $k\neq k'$. We shall make the reasonable assumption that noise 
terms for independent factors are independent, and make a further simplifying assumption 
that the release times of fake news associated with independent factors are likewise 
independent. 

A population of voters can be modelled by proposing a distribution over weight 
vectors $\{{\boldsymbol w}_n\}$. By randomly sampling from this distribution and 
computing, for each sample, which candidate is preferred, one can build up a 
population-level picture of voting patterns and investigate, in particular, the effect of 
fake news. In practice, such a distribution could be obtained by asking randomly chosen 
voters a series of questions in order to generate an approximation of their weight vectors. 
Moreover, large internet companies (such as search engines, social networks, and global retailers) 
possess 
data that is likely rich enough to reliably estimate these weights for individual users. This places such companies in a position of great responsibility. It should be evident 
that the knowledge of the weight vectors $\{{\boldsymbol w}_n\}$ is highly valuable in 
order to implement a targeted campaign. In particular, there is a concern that if such a targeted 
campaign were carried out by originators of fake news, democracy could be at risk. 
However, the simulation studies below support the notion that the mere knowledge 
that there might be fake news in circulation may be sufficient to eliminate the majority of the 
impact of fake news.

\section{Opinion polls in the microstructure model} 

We now present simulation results in the case where candidates are evaluated by three binary 
independent factors, in an election with two candidates. For concreteness, let us suppose 
that the three factors concern whether the candidate is liberal or conservative; whether 
the candidate is healthy or not; and whether the candidate is of good or bad character. 
For the purpose of this simulation, let us assume that the hidden characteristics of the two 
candidates, labelled here as $A$ and $B$, are such that $X^A_1 = 1$ (candidate $A$ 
is liberal) and $X^B_1 = -1$ (candidate $B$ is conservative); $X^A_2 = 1$ (candidate 
$A$ is in good health) and $X^B_2 = -1$ (candidate $B$ is in bad health); and 
$X^A_3 = 1$ (candidate $A$ has good character) and $X^B_3 = -1$ (candidate $B$ 
has bad character), the names and values being chosen for illustration.  Voters try to estimate these factor values based on 
the information available to them. 

We take the number of voters to be $N=1$ million, and we sample the weight vectors 
according to the following prescription:  we draw the weight vectors of $55 \, \%$ of 
voters from a three-dimensional normal distribution with standard deviation 0.4, 
Distribution $1$, centred at $(1, 1, 1)$, but restricted to the positive values for the 
second dimension and restricted to the interval $[0,1]$ for the third dimension. 
The 
truncation reflects the fact that voters are unlikely, for instance, to actively prefer 
candidates in ill health or with bad character. Members of this part of the population 
tend to prefer liberal, healthy candidates of good character. We then draw the weight 
vectors of the remaining $45 \, \%$ of the population from the same, truncated 
normal distribution (Distribution $2$), but centred at $(-1, 1, 0)$, representing voters 
that prefer a conservative, healthy candidate, but that are indifferent in relation to the 
candidate's character. The distributions are chosen purely for illustration. 

\begin{figure}[t!]
\begin{center} 
\includegraphics[angle=0,scale=0.88]{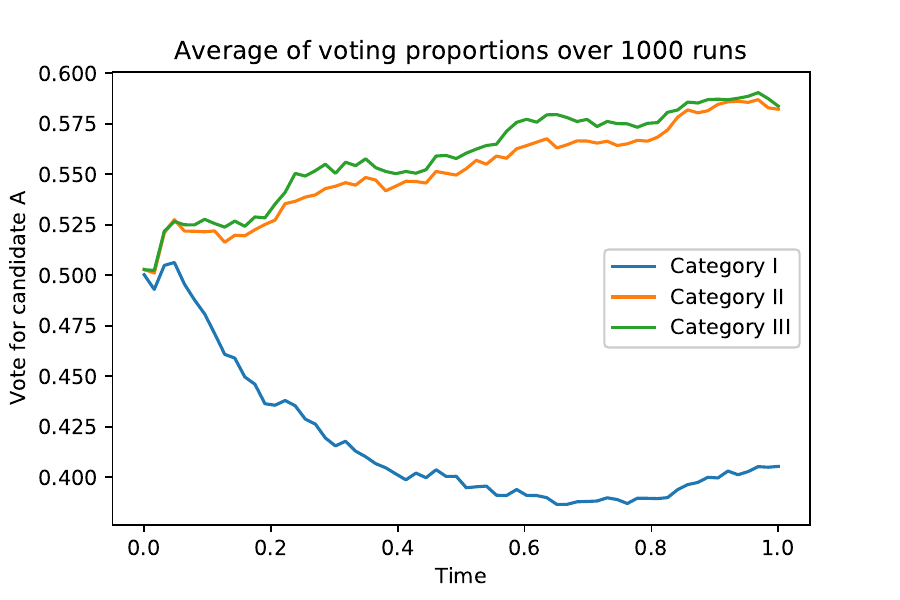}
\footnotesize
\caption{ 
\textit{Proportion of population voting for candidate $A$ as time passes}. 
These curves are taken as averages over 1,000 runs.  Category I voters are those that are unaware 
of the existence of fake news; category III voters are those who can fully eliminate fake 
news and thus represent the `correct' estimates. We see that fake news, which is 
designed to support candidate $B$, has the intended effect on average, making an 
election which candidate $A$ should win comfortably more competitive. These effects 
can be magnified or decreased by varying the parameters of the fake news terms in the 
information processes. We also recognize that Category II voters are successfully 
correcting for most of the effects of fake news. The parameters used in the simulation 
are: $\sigma=0.2$, $|\mu|=1.5$, $\alpha=4$, $p=0.5$, and $N=1$M. Waiting times 
between fake news releases are drawn at random for each run from an exponential 
distribution with  rate parameter  $4$. The code used for these simulations is available 
at {\tt github.com/dmmeier/Fake\_News}. 
\label{Fig_votes}}
\end{center}
\end{figure}

It is evident that candidate $A$, whose factor vector ${\boldsymbol X}^A = 
(1, 1, 1)$ coincides with the centre 
of Distribution $1$ representing $55 \%$ of the population, tends to win  the election in 
the absence of fake news.   
However, we now assume that in the run-up to the election various pieces of fake news 
are released that purport to show that the health and character of candidate $A$ are 
bad, that the health and character of candidate $B$ are good, that candidate 
$A$ is more conservative than they seem to be, and that candidate $B$ is more 
liberal. The intended effect of these elements of fake news is clear: shifting the perceived 
 vector of factors ${\boldsymbol X}^A$ of candidate $A$ out of the centre of Distribution $1$ while at the same 
time moving candidate $B$ towards it. Clearly, this increases the chances candidate $B$ has of 
winning the election. 

In order to isolate the effect of fake news in our simulations, we take the prior probabilities 
for all six random variables $\{X^A_k,X^B_k\}_{k=1,2,3}$ to be $0.5$ for the value $1$ 
and $0.5$ for the value $-1$. This means that the population starts out in a fully agnostic state. To model fake news we employ our earlier model in which the effect of a piece of 
fake news initially grows linearly in time but is then damped exponentially. 
In Figure \ref{Fig_votes} we show the average over 1,000 simulations, calculating voting proportions over time for each of the voter categories. 
The code used for these simulations is available at {\tt github.com/dmmeier/Fake\_News}. 

The results show that in the absence of fake news candidate $A$ on average wins the --
election quite handsomely -- this corresponds to the Category III curve. Fake news of the 
type constructed above, 
however, pushes the relative voting proportions of Category I voters, who are unaware 
of the existence of fake news, in favour of candidate $B$, as we had anticipated. 
What is perhaps most striking here is the curve showing estimates for Category II voters. These voters 
possess the knowledge that there may be fake news in circulation, but they do not know how many pieces of fake news have been released, or at what precise time.  That is, they only 
know the statistical distribution of 
$\{F_t\}$, or equivalently, the prior distribution $f_0(u)$. 
In spite of this, Category II voters are able to reduce the effects of fake news significantly by developing a statistical understanding 
of the nature of fake news and correcting their estimates accordingly. 
Furthermore, preliminary 
simulation studies suggest that their performance is robust against modest 
misspecifications of the prior distribution. 
 
\section{Discussion and outlook}

We have presented two approaches for the modelling of fake news in elections and 
referendums: one based on the idea of a 
representative voter, useful to obtain a qualitative understanding of the effects of fake news, and one 
based on the idea of an election microstructure, useful for practical implementation in 
concrete  scenarios. In both cases the results illustrate rather explicitly the impact of fake news in elections and referendums. We have demonstrated that by 
merely possessing the knowledge of the possibility that pieces of fake news might be 
in circulation, a diligent individual (Category II) is able to largely mitigate the effects of fake news. 

The models presented here invite further development in a number of directions. Our simulations, for instance, were based on random draws of waiting times for the release 
of fake news. However, in reality, a malicious individual trying to influence an election is 
likely to try to optimize release times to maximize impact (e.g., to maximize the 
chances of winning the election). It would be interesting to include 
such optimal release strategies in our models. Furthermore, as indicated above, in our 
simulations 
Category II voters are assumed to know the parameters of the fake news terms. This included, in particular, the value of the fake-news drift parameter $\mu$ and the damping rate $\alpha$. A natural extension of the model would allow for these parameters to be themselves random. Finally, the election microstructure approach could be developed further by allowing dependencies between the various factors, or by introducing several different information processes reflecting the news consumption preferences of different sections of society. 
These generalizations will open up  challenging but interesting new directions for research. 

At any rate, 
the performance of Category II voters, which significantly exceeded the expectations 
of the present authors, leads to a hopeful conclusion indeed: namely, by ensuring 
that members of the electorate are made aware of the possibility and the nature 
of fake news in the information they consume, policy makers may find success in 
countering the dark forces of fake news.

\section*{Acknowledgments}

The authors thank J.~Armstrong, A.~Macrina, and B.~K.~Meister 
for comments, and L.~P.~Hughston for conversations over the years on the 
information-based modelling framework.

\end{document}